\RequirePackage{amsmath}
\documentclass[10pt]{iopart}
\usepackage{iopams}

\usepackage{amsmath,amsfonts}
\usepackage{latexsym,amssymb}
\usepackage{amsthm}
\usepackage{bm}

\newcommand{\dd}{\mathrm{d}}

\begin{document}

\title[]{Comment on ``Potentials and fields of a charge set suddenly from rest into uniform motion'' by V. Hnizdo and G. Vaman}
\author{V. Onoochin}

\begin{abstract}
In this short Comment, the difference in the treatment of the gauge function presented in~\cite{VH1} and work of this author is analyzed. it is shown why some transformation of the gauge function made by Hnizdo and Vaman gives incorrect result.
\end{abstract}
 
 
In work titled  \lq{}Potentials and fields of a charge set suddenly from rest into uniform motion\rq{}~\cite{VH1},  the authors present the gauge function $\chi_{\rm C}$ which transforms the potentials defined in the Lorenz gauge into corresponding Coulomb gauge potentials. Such a gauge function for the potentials of uniformly moving charge is derived by one of the authors in~\cite{VH2}.  But a case of uniform motion of the charge is exceptional since the time variable can be eliminated from the expressions for the potentials by change $x'=x-vt$ and the system is reduced to the static one, where any application of the gauge transformation is not needed. Therefore, the demonstration of a gauge function that proves the transformation of time-varying potentials should be a significant step in the development of this topic.  
The author mention that \lq{} A recent eprint of V Onoochin \cite{On}, in which the case of a `charge that starts' is used, has provided a stimulus for doing it properly\rq{}. Since in the cited work, I prove that a gauge function for time--varying fields cannot exist in the general case, the existence of a gauge function (Eq.~(38) of~\cite{VH1}) should be considered as a refutation of the statement results presented in~\cite {On}. Therefore, it would be worthwhile to analyze the disagreements in this and the cited work.

In both works, the gauge function is introduced via its derivatives,
\begin{align}
\frac{\partial\bar{\chi}_{\rm C}(\bm{r},t)}{c\partial t}={\Phi}_{\rm L}(\bm{r},t)-{\Phi}_{\rm C}(\bm{r},t)\,, \quad
\boldsymbol{\nabla}\tilde\chi_{\rm C}(\bm{r},t)=\bm{A}_{\rm C}(\bm{r},t)-\bm{A}_{\rm L}(\bm{r},t)\,.
\label{chiC}
\end{align}
Here, notation of~\cite{VH1} is used. 

Although it seems obvious that the same gauge function enters into the above relations ($\bar{\chi}_{\rm C}=\tilde{\chi}_{\rm C}$), for rigour of consideration it would be expedient to prove this connection. The authors of~\cite{VH1} do this by transforming the second relation (chain of Eqs.~(29)--(33)). Since Eq.~(33) coincides with the second of Eqs.~(\ref{chiC}), the connection between the gauge functions is established. 

It should be noted a difference in approaches used in~\cite{On} and~\cite{VH1} to analyze the gauge function.

Despite transformations of $\chi_{\rm C}$ presented in~\cite{VH1} repeat some transformations given in~\cite{On} -- Eqs. (29)--(31) correspond to Eqs.~(2.3)--(2.5) of~\cite{On}, after deriving the expression for $\chi_{\rm C}$ via the  Lorenz--gauge scalar potential, 
\begin{align}
\chi_{\rm C}=-\frac{1}{4\pi c}\int \frac{\Dot{\Phi_{\rm L}}(\bm{r}',t)}{|\bm{r}-\bm{r}' |}\dd ^3r' \,,\label{Chi}
\end{align}
some more transformation are made by Hnizdo and Vaman to obtain the gauge function as an integral of the difference ${\Phi}_{\rm L}-{\Phi}_{\rm C}$ with respect to the time variable. But the chain of transformations contains certain error. 

The above integrand does not contain the scalar potential but its time derivative. Since a specific law of motion of the charge is considered, $\Dot{\Phi_{\rm L}}$ will be zero for some values of coordinates and time. This aspect is lost in Eq.~(33). Thus, in the chain of transformations (first three lines of Eq.~(33)), 
\begin{align}
\frac{\partial^2{\Phi}_{\rm L}(\bm{r}',t)}{c^2\partial t^2}\,\to\, 
\Box' {\Phi}_{\rm L}(\bm{r}',t)\,\to\, \rho (\bm{r}',t)\,\to\, \Phi_{\rm C}(\bm{r}',t)\,,
\end{align}
the boundary conditions for ${\Phi}_{\rm L}(\bm{r}',t)$ are ignored. Actually $\Dot{\Phi_{\rm L}}\neq 0$ in some closed region $0<r<ct$ -- the time-dependent Lorentz-gauge scalar potential spreads out in space as a spherical wave emitted from $r=0$ (p.$O$ in Fig.1 of~\cite{On}). But the transition to the wave equation for the Lorenz gauge scalar potential assumes that this non-zero scalar potential is defined in all space. Therefore, the gauge function is not zero (or constant) only in the region where the time derivative of ${\Phi}_{\rm L}$ is also non-zero. This puts a strong constraint on the final form of the gauge function that can be obtained from the integral. So instead of Eq.~(35) in~\cite{VH1} one should have
\begin{align}
\chi_{\rm C}(\bm{r},t)=c\int_{0}^t \dd t'\,[{\Phi}_{\rm L}(\bm{r},t')-{\Phi}_{\rm C}(\bm{r},t')]\Theta(ct-r)\,.
\label{chiC2}
\end{align}
where $t_0=0$ and $\Theta(\cdot)$ is the Heaviside step function.

As a result, the spatial derivative of $\chi_{\rm C}(\bm{r},t)$, or the difference in the vector potentials defined in the Coulomb and Lorenz gauges, should be equal to zero outside the region $0<r<ct$. It means, for example, that the term similar to $\dfrac{ct\bm{r}}{r^3}\Theta(r-ct)$ is absent. But this term provides the equivalence of the EM field calculated in both gauges in the region $r-ct>0$ since it compensates $\bm{\nabla}\dfrac{q}{r}\Theta(r{-}ct)$ there (the scalar potentials are presented by the expressions
\begin{align}
\Phi_{\rm L}(\bm{r},t)&=\frac{q\Theta(ct-r)}{\sqrt{(x-vt)^2+(y^2+z^2)/\gamma^2}} +\frac{q}{r}\,
\Theta(r-ct)\,,\label{Phi-L}\\
\Phi_{\rm C}(\bm{r},t)&=\frac{q\,\Theta(t)}{\sqrt{(x-vt)^2+(y^2+z^2)}}+\frac{q}{r}\,\Theta(-t)\,,
\end{align}
 and it is non-zero in that region). If $\bm{\nabla}\chi_C=0$ in this region, $\bm{A}_{\rm L}=\bm{A}_C$ and
 the electric field calculated in these gauges is
 \begin{align}
 {\bf E}_{\rm L}=-\bm{\nabla}\frac{q}{r}\Theta(r-ct)-\frac{\partial \bm{A}_{\rm L}}{c\partial t}\,\neq\,
 -\bm{\nabla}\frac{q\,\Theta(r-ct)}{\sqrt{(x-vt)^2+(y^2+z^2)}} -\frac{\partial \bm{A}_{\rm C}}{c\partial t}
={\bf E}_{\rm C}\,.
 \end{align}
Thus, the expressions for the electric fields are different and the found $\chi_{\rm C}$ does not perform its function, {\it i.e.} this gauge function does not exist.

In addition, one can compare the results of computation of $\bm{\nabla}\chi_{\rm C}=\bm{A}_{\rm C}-\bm{A}_{\rm L}$ made by Hnizdo and Vaman and those made in~\cite{On}. Since the integral (2.5) of~\cite{On} is complicated and it cannot be calculated in closed form for the potential~(\ref{Phi-L}), this integral is computed for $r=0$ (p.$O$ in Fig. 1 of~\cite{On}). The result, without the singular part due to derivative of the step--function, is
 \begin{align}
\partial_x\chi_{\rm C}(0,t)=-\frac{q}{(1-(v/c)^2)ct}\,.\label{I1}
 \end{align}
Coresponding time integration of difference in the potentials (Eqs.~(36) and (37) of~\cite{VH1}) in the region $0<r<ct$ and calculation of $\partial_x$ at $r=0$ gives
 \begin{align}
	\partial_x\chi_{\rm C}(0,t)&=\lim_{r\to 0}\frac{cq}{v}\left[\frac{1}{\sqrt{(r\cos\theta-vt)^2+r^2\sin^2\theta/\gamma^2}} - \frac{1}{\sqrt{(r\cos\theta-vt)^2+r^2\sin^2\theta}}\right]=
\nonumber\\
=&-\left(\frac{1}{\sqrt{1-(v/c)^2}}-1\right)\frac{cq}{v^2t}\,.\label{I2}	
\end{align}
where $\cos\theta=x/r$. Since a spatial derivative of the integral~(\ref{Chi}) is used in both works, the difference between direct computation of $\bm{\nabla}\chi_{\rm C}$ and some of its transformation and subsequent computation of the obtained result, Eqs.~(\ref{I1}) and~(\ref{I2}), should arise from the transformation of the gauge function in Eq.~(33) of~\cite{VH1}.
	
Finally it can be concluded that the function that transforms the Lorenz--gauge potentials into the Coulomb--gauge potentials such the charge motion, considered in the cited works, does not exist.


\section*{References}
\begin{thebibliography}{40}

\bibitem{VH1} Hnizdo V, Vaman G 2023 Potentials and fields of a charge set suddenly from rest into uniform motion. arXiv:2311.17652

\bibitem{VH2} Hnizdo V 2004  Potentials of a uniformly moving point charge in the Coulomb gauge {\it Eur.\,J.\,Phys.} {\bf 25} 351--60  (arXiv:physics/0307124)

\bibitem{On} Onoochin V 2023 Can [there] exist a function that transforms electromagnetic potentials from one to other gauge? arXiv:2305.15400

\end{thebibliography}
\end{document}